\magnification=\magstep1
\baselineskip=16pt
\hfuzz=6pt
\settabs 4 \columns
$ $

\vskip 1in

\centerline{\bf Quantum algorithms for topological and
geometric analysis of data}

\vskip 1cm 

\centerline{ Seth Lloyd$^1$, Silvano Garnerone$^2$, 
Paolo Zanardi$^3$}

\bigskip 
\centerline{1. Department of Mechanical Engineering, 
Research Lab for Electronics,}
 
\centerline{Massachusetts Institute of Technology, MIT 3-160,
Cambridge MA 02139, slloyd@mit.edu}

\centerline{2. Institute for Quantum Computing, University of Waterloo}

\centerline{3. Department of Physics and Astronomy, 
Center for Quantum Information Science \& Technology}

\centerline{ University of Southern California, Los Angeles, CA 90089-0484}

\vskip 1cm
{\bf Extracting useful information from
large data sets can be a daunting task.  Topological methods
for analyzing data sets provide a powerful technique for extracting
such information.  Persistent homology is a sophisticated tool
for identifying such topological features -- 
connected components, holes, or voids  -- and for determining how
such features persist as the data is viewed at different scales.
This paper provides quantum machine learning algorithms
for calculating Betti numbers in persistent homology, and
for finding eigenvectors and eigenvalues of the combinatorial
Laplacian.  The algorithms provide an exponential speedup
over the best currently known classical algorithms 
for topological data analysis.}

Human society is currently generating on the order of a billion billion
$(10^{18})$ bits of data a year.  Extracting useful information from
even a small subset of such a huge data set is difficult.  
A wide variety of big data
processing techniques have been developed to extract from large
data sets the hidden information in which one is actually interested.
Topological techniques for analyzing big data represent a sophisticated
and powerful tool [1-21]: by its very nature, topology reveals features
of the data that do not depend on how the data was sampled, how it
was represented, and how it was corrupted by noise.  Persistent
homology is a particularly useful topological technique that
analyzes the data to extract
topological features such as the number of connected components,
holes, voids, etc. (Betti numbers) of the underlying structure
from which the data was generated.  The length scale of analysis is then
varied to see whether those topological features persist 
at different scales.  A topological feature that persists over
many scales of analysis can be identified with a `true'
feature of the underlying structure.

Topological methods for analysis face challenges: a data
set that comes in the form of $n$ vectors 
possesses $2^n$ possible subsets that could contribute
to the topology.      Performing methods of algebraic topology
on simplicial complexes with $O{n \choose k +1}$  $k$-simplices
eventually requires matrix multiplication or diagonalization of matrices
of dimension $O{n \choose k+1}$.  For $k=O(1)$, such operations
require time ${\rm poly}(n^k)$.   Accordingly, when $k= O(n)$, 
matrix multiplication and diagonalization lead to 
problem solution scalings that go as ${\rm poly}(2^{n})$. 
A variety of mathematical methods 
have been developed to cope with the
resulting combinatorial explosion, notably mapping
the complex to a smaller complex with the same homology, and
then performing the matrix operations on the reduced complex [1-21].  
Even in such cases, the initial reduction must identify all
simplices in the original complex, and so can scale no better than 
linearly in the number of simplices.    
Consequently, even with only
a few hundred data points, creating the persistent
homology for Betti numbers at all orders $k$ is a difficult task.
In particular, the most efficient classical algorithms
for estimating Betti numbers at order $k$ (the number
of $k$-dimensional gaps, holes, etc.), have computational
complexity either exponential in $k$ or exponential in $n$
[8-11],  so that estimating Betti numbers to all orders
scales exponentially in $n$, and algorithms for diagonalizing
the combinatorial Laplacian at order $k$ have computational complexity
as $O( {n\choose k}^2) $, where 
$n$ is the number of vertices in the (possibly reduced) complex [11].
That is, the best classical
algorithms for estimating Betti numbers to all orders [10] and for diagonalizing
the full combinatorial Laplacian [11] grow exponentially in the number
of vertices in the complex.  This paper investigates
quantum algorithms for performing topological analysis of
large data sets.   We show that a quantum computer can estimate 
Betti numbers to all orders and to accuracy $\delta$ in time 
$O(n^3/\delta)$, and can find the eigenvectors and eigenvalues of
the combinatorial Laplacian to all orders and to accuracy
$\delta$ in time $O(n^5/\delta)$, in both cases reducing
a classical problem for which the best existing solutions
have exponential computational complexity, 
to a polynomial-time quantum problem.
Betti numbers can also be estimated by using a reduced,
or `witness' complex, that contains fewer points than the
original complex [1-10].    Applied to such witness complexes,
our method again yields an exponential speed-up in $n$. where
$n$ is now interpreted to be the number of points in the reduced
complex.
  
Recently, quantum mechanical techniques
have been proposed for machine learning and data analysis [22-31].
In particular, recent quantum machine algorithms [28-30] show that 
it is possible to obtain exponential speed-ups over the
best existing classical algorithms
for supervised and unsupervised learning.   
Such `big quantum data' algorithms use a quantum random access memory
(qRAM) [33-35] to
map an $N$ bit classical data set onto
the quantum amplitudes of a $\log_2 N$ qubit quantum state, 
an exponential compression over the classical representation.
The resulting state is then manipulated using quantum
information processing in time ${\rm poly}(\log_2 N)$
to reveal underlying features
of the data set.    That is, quantum computers that can
perform `quantum sampling' of data can perform certain
machine learning fasts exponentially faster than classical
computers performing classical sampling of data.
A discussion of computational complexity in quantum
machine learning can be found in [31]. 

Constructing a large scale qRAM to access
$N \sim 10^9 - 10^{12}$ pieces of data is a difficult
task.   By contrast, the topological and geometrical algorithms presented
here {\it do not} require a large-scale quantum RAM: a qRAM with
$O(n^2)$ bits suffices to store all pairwise distance information
between the points of our data set.  The algorithms presented here
obtain their exponential speed up over the best existing classical
algorithms not by having quantum access to a large data set, but
instead, by mapping a combinatorially large simplicial complex with
$O(2^n)$ simplices to a quantum state with $n$ qubits, 
and by using quantum information processing techniques such
as matrix inversion and diagonalization to perform 
topological and geometrical analysis exponentially faster
than classical algorithms.  Essentially, our quantum
algorithms operate by finding the eigenvectors and
eigenvalues of the combinatorial Laplacian [11].   
But diagonalizing a $2^n$ by $2^n$ sparse matrix using
a quantum computer takes time $O(n^2)$, compared
with time $O(2^{2n})$ on a classical computer.

The algorithms given here are related to quantum matrix inversion
algorithms [38].   The original matrix inversion algorithm [38]
yielded as solution a quantum state, and left open the question
of how to extract useful information from that state.    The topological
and geometric algorithms presented here answer that question:
the algorithms yield as output not quantum states but rather
topological invariants -- Betti numbers -- and do so in time
exponentially faster than the best existing classical algorithms.
The best classical algorithms for calculating the $k$'th Betti
number takes time $O(n^k)$, and estimating Betti numbers to
all orders to accuracty $\delta$ takes time at least $O(2^n \log(1/\delta))$ 
[8-11].
Exact calculation of Betti numbers for some types of topological
sets (algebraic varieties) is PSPACE hard [41].    By contrast, 
our algorithm provides approximate values of Betti numbers to all orders
and to accuracy $\delta$ in time $O(n^3/\delta)$:
although no polynomial classical algorithm for such approximate
evaluation of topological invariants is known,
the computational complexity of such approximation remains
an open problem.  We do not expect our quantum algorithms
to solve a PSPACE hard problem in polynomial time.

We summarize the comparison between the amount of resources required
by the classical and quantum algorithms in the following

\bigskip
\noindent{\it Table 1: Comparison of classical and quantum algorithms
for topological analysis}

\bigskip
\+ Procedural Steps && Classical Cost  & Quantum Cost\cr
\noindent $\hrulefill$

\+ Input pairwise distances, $n$ points && $O(n^2)$ bits & $O(n^2)$ bits \cr

\+ Construct simplicial complex&& $O(2^n)$ ops & 
$O(n^2)$ ops on $O(n)$ qubits\cr 

\+ Estimate all Betti numbers 
&& $O(2^{n} \log(1/\delta))$ ops & $O(n^3/\delta)$ quantum ops \cr

\+ Diagonalize combinatorial Laplacian && $O(2^{2n} \log(1/\delta))$ ops & 
$O(n^5/\delta)$ quantum ops\cr

\noindent $\hrulefill$

\noindent Here, $\delta$ is the multiplicative accuracy to which the 
Betti numbers and the eigenvalues of the combinatorial Laplacian  
are determined.  Note the tradeoff between the exponential quantum
speedup and accuracy: the quantum algorithms obtain an exponential
speedup over classical algorithms but provide an accuracy that
scales polynomially in $1/\delta$ rather than exponentially.
This feature arises from the nature of the quantum phase estimation/
matrix inversion algorithms, which obtain their exponential
speed up by estimating eigenvectors and eigenvalues
using a `pointer-variable' measurement interaction [38].
By contrast, classical algorithms need only keep $O(\log ({1/\delta}))$
bits of precision, but must perform $O(2^{2n})$ steps to diagonalize
$2^n \times 2^n$ sparse matrices.

The paper proceeds as follows.   First, we describe the classical `pipeline'
for constructing persistent homology, and show how the different sections
of this pipeline can be quantized.     After some preliminaries on
quantum state preparation and distance evaluation, we show how the
description of simplicial complexes can be mapped onto quantum states,
yielding an exponential compression in the quantum representation
of the data ($O(n)$ qubits) compared with the classical representation
($O(2^n)$ classical bits).     The simplex states are vectors in a
$2^n$ dimensional complex vector space over $n$ qubits.
We employ Grover's quantum search algorithm
to show how such simplex states can be constructed to a desired degree
of accuracy in $O(n^2)$ operations on $O(n)$ qubits.    

Having constructed simplex states,  in the next section 
we perform topological and geometrical analysis.
Many features of algebraic topology can be expressed in terms of 
eigenvectors and eigenvalues of the boundary map, which maps the 
vector space of $k$-simplices into the vector space of $k-1$ simplices. 
we show how quantum phase estimation can be used to
identify the kernels of the boundary map and to estimate their dimensions
to multiplicative accuracy $\delta$:
this procedure takes time $O(n^3/\delta)$ and allows us to estimate the 
Betti numbers of the simplicial complex.    By comparison, classical
algorithms for finding such kernels take time $O(2^{2n}  \log(1/\delta))$ for 
sparse matrix diagonalization.

Finally, we apply our quantum analysis to diagonalize the combinatorial
Laplacian, the discrete version of the ordinary continuous Laplacian.
The eigenvectors and eigenvalues of the combinatorial Laplacian
represent the fundamental modes (harmonic forms) and higher
order harmonics of the data.   As with the topological analysis,
his discretized geometric analysis takes time polynomial in 
the data set, $O(n^5)$ compared with $O(2^{2n})$ for the best
available classical algorithms.

We have phrased our algorithms in terms of the practical
problem of data analysis for the
simple reason that classical algorithms for persistent homology
and for finding features of the combinatorial Laplacian are
frequently used for data analysis.  We note, however, that 
our quantum algorithms can be regarded as purely mathematical constructions
providing exponential speed ups over the best available classical algorithms
for discrete topology and geometry.

\bigskip\noindent{\it The pipeline}

In this paper we present a quantum algorithm
for revealing topological features of the underlying structure:
dssentially, we quantize methods of persistent homology.
the algorithm operates by mapping vectors, simplices,
simplicial complexes, and collections of simplicial complexes
to quantum mechanical states, and reveals topology by
performing linear operations on those states.
The $2^n$ possible simplices of the simplicial complex are
mapped onto an $n$-qubit quantum state. Kernels of the
boundary map are then
found by conventional quantum computational techniques of 
eigenvector and eigenvalue analysis, matrix inversion, etc. 
The quantum analysis
reveals topological features of the data, and shows
how those features arise and persist when the scale
of analysis is varied.  The resulting quantum algorithms provide
an exponential speed-up over the best existing classical algorithms
for topological data analysis.  

In addition to constructing a quantum algorithm to reveal topological
features such as Betti numbers, we use the relationship between
algebraic topology and Hodge theory [10-21] to reveal geometrical
information about the data analyzed at different scales.   We
construct a quantum algorithm to identify the harmonic forms
of the data, together with the other eigenvalues and eigenvectors
of the combinatorial Laplacian -- the quantities that famously
allow one to `hear the shape of a drum' [32].   As with our quantum algorithm 
for finding Betti numbers, this geometric quantum algorithm is exponentially
faster than the corresponding classical algorithms.   
In particular, our quantum algorithm for finding all Betti numbers for
the persistent homology for simplicial complexes over $n$ points takes time
$O(n^3/\delta)$, and our algorithm for diagonalizing the combinatorial
Laplacian takes time $O(n^5/\delta)$, where $\delta$ is the multiplicative
accuracy to which Betti numbers and eigenvalues are determined.   
The best available classical algorithms to
perform these tasks at all orders $k$ take time $O(2^{2n}\log(1/\delta))$.

The advantage of big quantum data techniques is that they provide
exponential compression of the representation of the data.
The challenge is to see if -- and this is a big `if' --
it is still possible to process the highly-compressed quantum data 
to reveal the desired hidden structure that underlies
the original data set.  Here we show that quantum information
processing acting on large data sets encoded in a quantum
form can indeed reveal the persistent homology of the data set.

Classical algorithms for revealing persistent homology have
two steps (the `pipeline').  First, one processes the data
to allow the construction of a topological structure such
as a simplicial complex that approximates the hidden structure
from which the data was generated.  The details of the topological
structure depends on the scale at which data is grouped
together.  Second, one constructs topological invariants of that 
structure and analyzes how those invariants behave as a function
of the grouping scale.  As above, topological invariants that persist 
over a wide range of scales are identified as features
of the underlying hidden structure.

The quantum `pipeline' for persistent homology also has two
steps.  First, one accesses the data in quantum parallel to 
construct quantum states that encode the desired topological
structure: if the structure is a simplicial complex, for
example, one constructs quantum states that are uniform
superposition of descriptions of the simplices in the complex.
second, one uses the ability of quantum computing to reveal
the ranks of linear maps to construct the topological invariants
of the structure.  The steps of the quantum pipeline
are now described in more detail.

For example, a classical data set
containing $2^{300}$ bits -- the total amount of information
that could be stored within the observable universe
if every single elementary particle registers the maximum
amount of information allowed by the laws of physics --
could be encoded in just $300$ quantum bits.   

\bigskip\noindent{\it Quantum preliminaries: state
preparation and distance evaluation}

Topological analysis of the data requires distances
between data points.  Assume that the data set contains
$n$ points together with the $n(n-1)/2$ distances between them. 
the data is stored in quantum random access memory or qRAM
[33-35], so that the algorithm can access the data in quantum parallel.   
The essential
feature of a quantum RAM is that it preserves quantum
coherence: the qRAM maps a quantum superposition of inputs 
$\sum_j \alpha_j |j\rangle |0\rangle$ to
a quantum superposition of outputs $\sum_j \alpha_j |j\rangle |v_j\rangle$.
Note that a quantum RAM is potentially significantly
easier to construct than a full-blown quantum computer. 
the storage medium of a quantum RAM can be essentially classical: indeed,
a single photon reflected off a compact disk encodes in its quantum
state all the bits of information stored in the mirrors on the disk.
in addition to a classical storage medium such as a CD, a qRAM 
contains quantum switches that can be opened in quantum
superposition to access that information in quantum parallel.
each call to an $N$-bit qRAM requires $\log_2 N$ 
quantum operations.
Quantum RAMS have been designed, and prototypes have
been constructed [33-35].
In contrast to other big quantum data algorithms [28-30], 
the size of the qRAM required to perform topological and
geometric analysis is relatively small: because the computational
complexity of classical
algorithms for persistent homology scales as $O(2^{2n})$, while
the quantum algorithms require only $O(n^2)$ bits worth of qRAM,
a significant quantum advantage could be obtained by a qRAM with
hundreds to thousands of bits.

As an alternative to being presented with the pre-calculated
distances, the data set could consist 
of $n$ $d$-dimensional vectors $\{ \vec v_j \}$ over
the complex numbers, and we can use the qRAM to 
construct the distances $|\vec v_i - \vec v_j|$ between 
the $i$th and $j$th vectors [28].  Finally, the distances
can be presented as the output of a quantum computation.
In all cases, our quantum algorithms for topological
and geometric analysis
operate by accessing the distances in quantum parallel.
Big quantum data analysis works by mapping each vector
$\vec v_j$ to a quantum state $|v_j\rangle \in C^{d}$,
and the entire database to a quantum state
$(1/\sqrt n) \sum_j |j\rangle |v_j\rangle \in C^n\otimes C^d$.
A quantum RAM can be queried in quantum parallel:
given an input state $|j\rangle|0\rangle$, it produces
the output state $|j\rangle|v_j\rangle $,
where $|v_j\rangle$ is normalized quantum state proportional
to the vector $\vec v_j$. 
Such a quantum state can be encoded using $O(\log_2(nd))$
quantum bits.  
and $|\vec v_j|$ is the norm of the vector.  

If we have not been given the $n(n-1)/2$ distances directly in 
quantum RAM,
the next ingredient of the quantum algorithm is the ability to
evaluate inner products and distances between vectors. 
In [18,28-30] it is shown how the access to vectors in quantum
superposition: the ability to create the quantum states
corresponding to the vectors translates into the ability
to estimate $|\vec v_i - \vec v_j|^2 
= 2 - {\vec v}_i^\dagger \vec v_j 
- {\vec v}_j^\dagger \vec v_i$.  
That is, we can construct a quantum circuit that takes
as input the state $|i\rangle|j\rangle|0\rangle$ and
produces as output the state 
$|i\rangle|j\rangle| |\vec v_i - \vec v_j|^2 \rangle$,
where the third register contains an estimate of
the distance between $\vec v_i$ and  $\vec v_j$. 
To estimate the distance to accuracy $\delta$
takes $O(\delta^{-1})$ quantum memory calls 
and $O(\delta^{-1} \log_2(nd))$ quantum operations.
As with the quantum random access memory, the circuit
to evaluate distances operates in quantum parallel.

\bigskip\noindent{\it Step 1: Constructing a simplicial complex}

Classical persistent homology algorithms
use the access to data and distances to construct a topological
structure -- typically a simplicial complex -- that corresponds
to the hidden structure whose topology one wishes to reveal.
in the quantum algorithm, we use the ability to access data and to estimate
distances in quantum parallel to construct quantum 
states that encode the simplicial complex.  

Each simplex in the complex consists of a fully connected set of vertices:
a $k$-simplex $s_k$ consists of $k+1$ vertices
$j_0 j_1 \ldots j_k$ (listed in ascending
order, $j_0 < j_1 < \ldots < j_k$) together 
with the $k(k+1)/2$ edges connecting
each vertex to all the other vertices in the simplex.
Encode a $k$-simplex $s_k$ as a string of $n$ bits, with $k+1$
$1$s at locations $j_0 j_1 \ldots j_k$ designating the vertices in the simplex.
removing the $\ell$th vertex and its associated edges from
a $k$ simplex yields a $k-1$ simplex. 
The $k+1$ simplices $s_{k-1}(\ell)$ with vertices 
$j_0 \ldots {\hat j}_\ell \ldots j_k$ obtained
by removing the $\ell$th vertex $j_\ell$ from $s_k$ form the boundary
of the original simplex.
The number of potential simplices in a simplicial complex
is equal to $2^n$, the number of possible subsets of the $n$ points
in the graph.  That is, every member of the power set is a potential
simplex.  If $n$ is large, the resulting combinatorial explosion
means that identifying large simplices can be difficult.

There are a variety of ways to construct such a simplicial
complex classically.  One can construct a cover of the complex,
and then form the simplicial complex from the points and intersections
of that cover.  The  \v Cech complex, for example, is constructed from
the intersections of a cover of $\epsilon$ balls centered at each
data point.  Alternatively, one can construct a graphical version
of the simplex: the popular Vietoris-Rips complex $V_\epsilon$
contains as its simplices subsets of points that are
all within $\epsilon$ of each other.   The form of the simplicial
complex $S^\epsilon$ depends on the scale $\epsilon$ at which its
points are grouped together: persistent homology investigates
how topological invariants of the simplicial complex 
depend on the scale $\epsilon$.

The collection of simplicial complexes $\{ S^\epsilon \}$ for
different values of the grouping scale $\epsilon$ is
called a filtration.  Note that if a simplex
belongs the complex $S^\epsilon$, then it also belongs
to $S^{\epsilon'}$, $\epsilon' > \epsilon$.  That is, 
the filtration consists of a sequence of nested simplicial
omplexes.  
For any simplex $s$, define $\epsilon(s)$ to
be the scale at which the simplex enters the complex $S_\epsilon$.

In the quantum case, just
as in the classical, there are a variety of methods for 
constructing a simplicial complex from the data.  Here we show
how to construct quantum states that correspond to the Vietoris-Rips
omplex.   The \v Cech complex can be constructed in a similar fashion
[37]. 
represent a simplicial complex $S^\epsilon$ over $n$ vertices
as a graph: the vertices are the vectors in the data set,
and the edges connect vectors that are less than $\epsilon$ apart.  
as above, encode simplices as quantum states over $n$ qubits
with $1$s at the positions of the vertices.  We designate the 
$k$-simplex $s_k$ by the $n$-qubit basis vector
$|s_k\rangle \in C^{2^n}$. 
Denote the ${n \choose k+1}$ dimensional Hilbert space corresponding
to $k$ simplices by $W_k$.
Let ${\cal H}_{k}^{\epsilon}$ be the subspace
of $W_k$ spanned by $|s_k\rangle$ where
$s_k \in S_{k}^{\epsilon}$, the set of $k$ simplices
in $S^{\epsilon}$.  The full simplex space at scale $\epsilon$
is defined to be ${\cal H}^\epsilon = \oplus_k {\cal H}_{k}^{\epsilon}$.
our ability to evaluate distances translates
onto the ability to apply the projector $P_{k}^{\epsilon}$ that
projects onto the $k$-simplex space ${\cal H}_{k}^{\epsilon}$ and the projector
$P^\epsilon$ that projects onto the full simplex space ${\cal H}^\epsilon$.

We now use Grover's algorithm 
to construct the $k$-simplex state 
$$|\psi\rangle_{k}^{\epsilon} = {1 \over \sqrt{|S_k^\epsilon|} } 
\sum_{s_k \in S_k^\epsilon} |s_k\rangle,\eqno(1)$$
where as above $S_{k}^{\epsilon}$ is the set of $k$-simplices in the complex
at scale $\epsilon$.
That is, $|\psi\rangle_{k}^{\epsilon}$ is the uniform superposition
of the quantum states corresponding to $k$-simplices in
the complex.  We employ the multi-solution version of Grover's
algorithm: for each simplex $s_k$ we can verify whether
$s_k\in S_k^\epsilon$ in $O(k^2)$ steps.   That is, we
caan implement a membership function $f_k^\epsilon(s_k) = 1$
of $s_k \in S_k^\epsilon$ in $O(k^2)$ steps.    Grover's algorithm [31] then 
allows us to construct the $k$-simplex state (1).     
the construction of the $k$-simplex state
via Grover's algorithm reveals the number of $k$-simplices
$|S_k^\epsilon| = {\rm dim} H_k^\epsilon$ in the complex
at scale $\epsilon$, and takes time 
$O( n^2(\zeta_k^\epsilon)^{-1/2}$, 
where $\zeta_k^\epsilon = |S_{k}^{\epsilon}|/{n\choose k +1}$ is the
fraction of possible $k$-simplices that are actually
in the complex at scale $\epsilon$.     
When this fraction is too small, the quantum
search procedure will fail to find the simplices.
For $k<<n$, we have ${n \choose k+1} =  O(n^{k+1}/k\!)$, and 
$\zeta_k^\epsilon$ is only polynomially small in $n$.
By contrast, for $k\approx n$, $\zeta_k^\epsilon$ can
be exponentially small in $n$: if only an exponentially small
set of possible simplices actually lie in the complex,
Quantum search will fail to find them.  As $\epsilon$ increases,
however, more and more simplices enter into the complex.  Above
some value of $\epsilon$, quantum search will succeed in
constructing the simplex state.   When $\epsilon$ becomes larger than
the maximum distance between vectors, all simplices are in the
complex.  

Below, it will prove useful to have, in addition to the simplex
state $|\psi\rangle^\epsilon_k$, the state 
$\rho^\epsilon_k = (1/|S_k^\epsilon|)
\sum_{s_k \in S_k^\epsilon} |s_k\rangle\langle s_k|$,
which is the uniform mixture of all $k$-simplex states in
the complex at grouping scale $\epsilon$.   $\rho^\epsilon_k$
can be constructed in a straightforward fashion from the
simplex state $|\psi\rangle^\epsilon_k$ by adding
an ancilla and copying the simplex label to construct
the state 
${1 \over \sqrt{|S_k^\epsilon|} }
\sum_{s_k \in S_k^\epsilon} |s_k\rangle\otimes|s_k\rangle$.
tracing out the ancilla then yields the desired uniform
mixture over all $k$-simplices.

To elucidate the construction of the $k$-simplex states (1), we
look more closely into the implementation of Grover's algorithm to
understand when it succeeds in constructing the $k$-simplex state,
and how it fails. 
Start from a superposition $n^{-1/2} \sum_k |k\rangle$ over
all values of $k$.
Performing simplex construction
in parallel via Grover's algorithm with the membership 
function $f_k^\epsilon$ 
yields the full simplex state at scale $\epsilon$:
$$|\Psi\rangle^\epsilon
= {1\over \sqrt n} \sum_k |k\rangle |\psi\rangle_k^\epsilon.\eqno(2)$$
By adding ancillae as above, we can also construct the uniform mixture over
all values of $k$ and and all $k$-simplices: 
$\rho^\epsilon = (1/n) \sum_k |k\rangle\langle k|
\otimes \rho_k^\epsilon$.

More precisely,
if we run the quantum search procedure for a time $\zeta^{-1/2}$, 
then we will obtain the state
$$|\Psi\rangle^{\epsilon}_{\zeta} 
= {1\over \sqrt n} \bigg( \sum_{k:\zeta_k^\epsilon \geq \zeta}
|k\rangle|\psi\rangle_k^\epsilon + 
\sum_{k:\zeta_k^\epsilon < \zeta}
|k\rangle|0\rangle \bigg) \eqno(3) $$ 
which contains the simplex
states $|\psi\rangle_k^\epsilon$ for which
$\zeta_k^\epsilon \geq \zeta$ and which
returns a null result $|0\rangle$ for
the simplex states for which 
$\zeta_k^\epsilon < \zeta$.     
For small $\epsilon$ -- where only a small fraction of all possible
simplices lie within the complex -- and fixed $\zeta$, the simplex state 
$|\Psi\rangle^{\epsilon}_{\zeta}$ will contain the actual
simplex states $|\psi\rangle_{k}^{\epsilon}$ only for
small $k$.   
As $\epsilon$ becomes larger and larger, higher
and higher $k$-simplex states enter the filtration and
$|\Psi\rangle^{\epsilon}_{\zeta}$ will contain more and more of
the $k$-simplex states.  

Constructing the simplex state in quantum parallel
at $m$ different grouping scales $\epsilon_i$ yields the filtration
state
$$|\Phi\rangle_\zeta = 
{1\over \sqrt{mn}} \sum_i |\epsilon_i\rangle |\Psi\rangle^{\epsilon}_{\zeta}.
 \eqno(4)$$
The filtration state $|\Phi\rangle_\zeta$ contains the entire filtration
of the simplicial complex in quantum superposition.
The quantum filtration state contains exponentially fewer quantum bits 
than the number of classical bits required to describe the classical
filtration of the complex: $\log m$ qubits are required to register
the grouping scale $\epsilon$, and $n$ 
qubits are required to label the simplices.
$|\Phi\rangle_\zeta$ takes time $O(\zeta^{-1/2} n^2 \log(m))$
to construct.  By contrast, a classical description of the filtration of
the simplicial complex requires $O(2^n)$ bits. 


This section showed that we can represent the 
full filtration of the simplicial complex
in quantum mechanical form using exponentially fewer bits than
are required classically.  Indeed, the quantum search method for
constructing simplicial states works best when $\zeta_{k}^{\epsilon}$ is
not too small, so that a substantial fraction of simplices that could
be in the complex are actually in the complex.   But this regime is
exactly the regime where the classical algorithms require an exponentially
large amount of memory
space $O(\zeta_{k}^{\epsilon} {n\choose k+1})$ bits merely to record
which simplices are in the complex.   Now we show how to
act on this quantum mechanical representation of the 
filtration to reveal persistent homology.

\bigskip\noindent{\it Step 2: Topological analysis}

In the analysis of persistent homology, having constructed
a simplicial complex $S^{\epsilon}$ at scale $\epsilon$, one
analyzes its topological properties.  The first step of the
quantum version of
the topological analysis is to map $k$-simplices $s_k$
to basis vectors $|s_k\rangle$
in a vector space $W_k$ with dimension ${\rm dim} W_k = {n \choose k+1}$.  
As above, let ${\cal H}_{k}^{\epsilon}$ 
be the space $\in W_k$ spanned by vectors
corresponding to $k$-simplices in the complex at level $\epsilon$.
We identify the vector space ${\cal H}_{k}^{\epsilon}$ with
the abelian group $C_k$ (the $k$'th chain group)
under addition of vectors in the space.
Let $j_0\ldots j_k$ be the vertices of $s_k$.
Define the {\it boundary map} $\partial_k$ from
$W_k$ to $W_{k-1}$ by
$$ \partial_k |s_k\rangle
= \sum_\ell (-1)^\ell |s_{k-1}(\ell)\rangle\eqno(5)$$
where as above $s_{k-1}(\ell)$ is the $k-1$ simplex on the boundary of $s_k$
with vertices $j_0 \ldots {\hat j}_\ell \ldots j_k$
obtained by omitting the $\ell$'th vertex $j_\ell$ from $s_k$.
The boundary map maps each simplex to the oriented sum of its
boundary simplices.
$\partial_k$ is a ${ n\choose k} \times {n\choose k+1}$ matrix
with $n-k$ non-zero entries $\pm 1$ in each row and $k+1$ non-zero entries
$\pm 1$ per column. 
Note that $\partial_k \partial_{k+1} = 0$: the boundary
of a boundary is zero.   As defined, $\partial_k$ acts on the space
of all $k$-simplices.   We can also define the boundary map
restricted to operate from
${\cal H}_{k}^{\epsilon}$ to ${\cal H}_{k-1}^{\epsilon}$,
to be $\tilde\partial_k \equiv P_{k-1}^\epsilon \partial_k P_k^\epsilon$.

The $k$'th homology group ${\rm {\bf H}}_k$ is the quotient 
group, ${\rm Ker}~\partial_{k} / {\rm Image}_{k+1} \partial_{k+1}$,
the kernel of $\partial_k$ divided by
the image of $\partial_{k+1}$ acting on ${\cal H}_{k+1}$ at 
grouping scale $\epsilon$.  
The $k$th Betti number $\beta_k$ is equal to the dimension
of ${\rm {\bf H}}_k$, which in turn is equal to
the dimension of the kernel 
of $\partial_k$ minus the dimension of the image 
of $\partial_{k+1}$.

The strategy that we use to identify persistent topological
features starts by identifying the singular values and singular vectors of the
boundary map.  Connected components, holes, voids, etc., correspond to
structures -- chains of simplices -- that have no boundary, but that
are not themselves a boundary.  That is, we are looking for the set
of states that lie within the kernel of $\partial_k$, but that do
not lie within the image of $\partial_{k+1}$ acting on the vector
space of $k+1$ simplices 
${\cal H}_{k+1}^\epsilon$.
The ability to decompose arbitrary vectors in
$W_k$ in terms of these kernels and images 
allows us to identify Betti numbers at different grouping scales $\epsilon$.


The quantum phase algorithm [34-35] allows one to decompose states in
terms of the eigenvectors of an Hermitian matrix and to find
the associated eigenvalues.   
Once the $k$-simplex states
$|\psi\rangle_{k\epsilon}$ have been constructed, the quantum
phase algorithm allows one to decompose those states in terms of eigenvectors
and eigenvalues of the boundary map.  
The boundary map is not Hermitian.  We can embed the boundary map
into a Hermitian matrix $B_k$ defined by
$$ B_k = \pmatrix{ 0 &  \partial_k \cr \partial_k^\dagger & 0\cr}.\eqno(6)$$
$B_k$ is $n$-sparse and acts on the space $W_{k-1} \oplus W_k$. 

Now apply the quantum phase algorithm to $B_k$ starting with initial
state the uniform mixture of $k$-simplex states $\rho_{k}^{\epsilon}$.  The
quantum phase algorithm decomposes $\rho_k^\epsilon$
into the eigenstates of $B_k$ and reveals the corresponding
eigenvalue to accuracy $\delta$, taking time $O(n^3\delta^{-1})$ 
[33].
If one then measures the eigenvalue register, one obtains a
particular eigenvalue with a probability equal to the
dimension of the corresponding eigenspace, divided by the dimension
of ${\cal H}_k^\epsilon$.   That is, in addition
to the eigenvalues and eigenvectors, the quantum phase algorithm
reveals the dimension
of the corresponding eigenspaces.
In particular, the quantum phase algorithm can be used to
project $|\psi\rangle_{k}^{\epsilon}$ onto the kernel 
of $B_k$, which is equivalent to projecting it onto 
to the kernel of $\partial_k$.   The projection succeeds with
probability $\eta_{k}^{\epsilon}
= {\rm dim}~ ({\rm Ker}~\partial_k) / |S_k^\epsilon|$,
where as above $|S_{k}^{\epsilon}| = {\rm dim}~{\cal H}_k^\epsilon$,
the dimension of the space of $k$-simplices at scale $\epsilon$,
thereby revealing the dimension of that kernel. This process
allows us to reconstruct the
$k$'th Betti number 
$$\beta_k = {\rm dim} ~ {\rm Ker}~ \partial_k 
- {\rm dim}~ {\rm Im}~\partial_{k+1} =
{\rm dim}~ {\rm Ker}~\partial_k
+ {\rm dim}~ {\rm Ker}~\partial_{k+1} - {\rm dim}~{\cal H}^\epsilon_{k+1}
\eqno(7)$$
to accuracy $\delta$ in time 
$$\max( O(n^3/\eta_k^\epsilon \delta), O(n^2 (\zeta_{k}^{\epsilon})^{-1/2}). 
\eqno(8)$$ 
That is, the computational complexity is dominated either by
the quantum phase algorithm and the probability that one identifies
an eigenstate in the kernel of $B_k$ (first term in the $\max$),
or by the construction of the simplex state (second term in the $\max$). 
The algorithm uses $O(n^2)$ bits of quantum random access memory to
store the distances, and requires $O(n)$ qubits of `core' memory
in the quantum computer.
By contrast, when performing the topological analysis on a classical
computer, merely to write down the set
of simplices in the filtered simplicial complex at level $k$ takes classical 
space up to $O{n \choose k_1}$ bits, as $\epsilon$ increases,
and to write down the set of simplices
at all levels $k$ takes space $O(2^{2n})$. 
Constructing the $k$th Betti number using
sparse matrix diagonalization or rank finding takes time 
time $O\big( {n\choose k}^2 \big)$ which goes as
$O(2^{2n})$ for $k = O(n)$.   
The quantum phase algorithm allows us to construct
the Betti numbers $\beta_k$ for the simplicial complex for all $k$
exponentially faster than classical algorithms.     

The quantum algorithm for constructing Betti numbers can be implemented
in quantum parallel, yielding a quantum state that contains
the Betti numbers at each scale $\epsilon$ in quantum superposition.
Once the information about Betti
numbers at different scales has been presented in quantum mechanical
form, we can also use the tools
of quantum information processing -- matrix inversion [33], quantum
Fourier transforms, etc. -- to reveal correlations and patterns
between Betti numbers, simplex states, and grouping scales. 

\bigskip\noindent{\it 3. Quantum algorithm for finding eigenvalues
and eigenvectors of the combinatorial Laplacian}

In addition to revealing topological information -- the Betti
numbers -- at different length scales, the quantum phase
algorithm can be used to reveal geometric information.   We
now present a quantum algorithm for decomposing the simplicial
complex in terms of eigenvectors and eigenvalues of the combinatorial
Laplacian.   This decomposition contains useful geometric
information about the data [10-21], such as harmonic forms.
It also represents an alternative way of constructing the 
Betti numbers.

The first step of the construction is to restrict the
boundary map to operate only on the simplicial subspace
${\cal H}^\epsilon$ at scale $\epsilon$.   As noted above, the ability
to evaluate distances in quantum parallel translates
into the ability to apply the projector 
$P^\epsilon$ that projects onto ${\cal H}^\epsilon$.  Because it
involves evaluating the $k(k+1)/2$ distances in each $k$-simplex in 
quantum parallel,
the application of this projector takes time $O(n^2)$.
Define the full Hermitian boundary map 
$$B^\epsilon = 
 \pmatrix{ 0 &  \tilde\partial_1 & 0  &&&  \cr 
\tilde\partial_1^\dagger & 0 & \tilde\partial_2 &&\ldots& \cr
0 & \tilde\partial_2^\dagger & 0 &&& \cr
&&& \ldots &&\cr
&&&0 &\tilde\partial_{n-1} &0\cr
&\ldots &&\tilde\partial_{n-1}^\dagger & 0 & \tilde\partial_n\cr
&&& 0&\tilde\partial_n^\dagger & 0\cr },\eqno(9)$$
where as above $\tilde\partial_k = P^\epsilon \partial_k P^\epsilon$ is
the boundary map confined to the simplicial subspace ${\cal H}^\epsilon$.
Note that $B^\epsilon$ is $n$-sparse: there are at most $n$ entries
in each row.
Because $\tilde\partial_k\tilde\partial_{k+1} =0$, we have
$${B^\epsilon}^2 =  
\pmatrix{ 
\tilde\partial_1 \tilde\partial_1^\dagger & 0 &0& \cr
0& \tilde\partial_1^\dagger \tilde\partial_1 
+ \tilde\partial_2  \tilde\partial_2^\dagger &0&\ldots\cr
0&0& \tilde\partial_2^\dagger \tilde\partial_2 + 
\tilde\partial_3 \tilde \partial_3^\dagger &&\cr
&&\ldots && \cr
&\ldots&& \tilde\partial_{n-1}^\dagger\tilde\partial_{n-1} + 
\tilde \partial_n \tilde\partial_n^\dagger&0\cr
&&&0& \tilde\partial_n^\dagger \tilde\partial_n\cr} .\eqno(10)$$ 
That is, ${B^\epsilon}^2 = 
\Delta_0 \oplus \Delta_1 \oplus \ldots \oplus \Delta_n$, where 
$\Delta_k = ~ \tilde\partial_k^\dagger \tilde\partial_k  
+  \tilde\partial_{k+1} \tilde\partial_{k+1}^\dagger $  
 is the combinatorial Laplacian of the $k$th simplicial complex [20-21].
Because ${B^\epsilon}^2$ is the sum of the combinatorial Laplacians,
$B^\epsilon$ is sometimes called the `Dirac operator,' since the original Dirac
operator was the square root of the Laplacian.
Hodge theory [10-21] implies that the $k$'th homology group $
{\rm {\bf H}}_k = 
 {\rm Ker}~ \tilde\partial_{k} / {\rm Image}_{k+1} \tilde\partial_{k+1} \cong
{\rm Ker}~ \Delta_k$.  As above, the dimension of this
kernel is the $k$'th Betti number.

Now apply the quantum phase algorithm to $B^\epsilon$ starting
from the uniform mixture of simplices $\rho^\epsilon$.   The quantum 
phase algorithm decomposes the simplex state into the eigenvectors
of the combinatorial Laplacian, and identifies the corresponding
eigenvalues.   The probability of yielding a particular eigenvalue
is proportional to the dimension of the corresponding eigenspace.
As above, classical algorithms for finding the eigenvalues and
eigenvectors of the combinatorial Laplacians
$\Delta_k$, and calculating the dimension of the eigenspaces takes 
$O( {n\choose k}^2 ) \sim O(2^{2n})$ 
computational steps using sparse matrix diagonalization via
Gaussian elimination or the Lanczos algorithm.   On a quantum computer,
however, the quantum phase algorithm [39-40] can project the simplex
states $|\psi\rangle_{k}^{\epsilon}$ onto the eigenspaces of 
the Dirac operator
$B^\epsilon$ and find corresponding eigenvalues to
accuracy $\delta$ in time $O(n^5\delta^{-1} \zeta^{-1/2})$, where
as above $\zeta$ is the accuracy to which we choose to construct
the simplex state.  
The algorithm also identifies the dimension of the
eigenspaces in time $O(n^5\delta^{-1} \zeta^{-1/2} \eta_\ell^{-1/2})$,
where $\eta_\ell$ is equal to the dimension $d_\ell$ of the $\ell$th eigenspace
divided by $|S|_k$, the dimension of the $k$-simplex space.
The $k$th Betti number $\beta_k$ is equal to the dimension of
the kernel of $\Delta_k$.  The additional
factor of $n^2$ compared with the algorithm for finding the
Betti numbers alone comes from the need to project onto the simplicial
subspace each stage of the computation -- i.e., to apply the $P^\epsilon$.
The additional computation allows us to construct the full decomposition
of the simplicial complex in terms of eigenvectors and eigenvalues
of the combinatorial Laplacian, yielding useful geometric information
such as harmonic forms.
Monitoring how the eigenvalues and eigenspaces
of the combinatorial Laplacian change
as $\epsilon$ changes provides geometric information about
how various topological features such as connected components,
holes, and voids come into existence and disappear as the
grouping scale changes.

\bigskip\noindent{\it Discussion}

This paper extended methods of quantum machine learning 
to topological data analysis.   
Homology is a powerful topological tool.  
The representatives of the homology classes for different
$k$ define the connected components of the simplicial
complex, holes, voids, etc.  The Betti numbers count
the number of connected components, holes, voids, etc. 
Varying the simplicial scale $\epsilon$ and plotting how
Betti numbers change as function of $\epsilon$ reveals 
how topological features come into existence and go away
as the data is analyzed at different length scales.
Our algorithm also reveals how the structure of the eigenspaces
and eigenvalues of the combinatorial Laplacian changes as a function
of $\epsilon$.   This `persistent geometry' reveals features of
the data such as rate of change of harmonic forms over different
simplicial scales.

The underlying methods of our quantum algorithms
are similar to those in other big quantum data algorithms [17-19].
The primary difference between the topological and geometrical
algorithms presented here, and algorithms for, e.g., constructing
clusters [17], principal components [18], and support vector
machines [19], is that our topological algorithms require only
a small quantum random access memory of size $O(n^2)$.  Consequently,
even when the full qRAM resources are included in the accounting of 
the computational complexity of the algorithms, the topological
algorithms require only an amount of computational resources polynomical
in the number of data points, while the best existing classical
algorithms for answering the same questions require exponential resources.

To recapitulate:

\noindent (1) The classical data is mapped via a quantum random access
memory into a tensor product quantum state.

\noindent (2)  The quantum data is processed
using standard techniques of quantum computation:
distances between vectors are evaluated, simplices of
neighboring vectors are identified, and a simplicial
complex is constructed.  The simplicial complex
depends on the grouping scale $\epsilon$.  We construct
a quantum state that represents the filtration of the
complex -- the set of simplicial complexes, related
by inclusion, for different $\epsilon$.
This quantum state contains exponentially fewer qubits
than the number of bits required to describe the classical
filtration of the complex.

\noindent (3) Now construct homology in quantum parallel.
Perform the boundary map to associate each
simplex in the filtration with its boundary.
Standard techniques of quantum information
processing then allow one to identify the
dimensions of the kernel and image of the boundary
map for each $k$.  This in turn allows us to
calculate Betti numbers for each $k$ and
each scale $\epsilon$.

\noindent (4) Use the quantum phase algorithm to calculate the
eigenvalues and to construct the eigenspaces 
of the combinatorial Laplacian at each scale.
This construction gives us geometric information
about the data set.

Classical algorithms for performing the full persistent homology over a space
with $n$ vectors over all scales $k$ take time $O( 2^{2n} )$:
there are $2^n$ possible simplices, and evaluating kernels and
images of the boundary map via Guassian elimination for sparse
matrices takes time
that goes as the square of the dimension of the space of simplices.
By contrast, the quantum algorithm for constructing the Betti numbers
in quantum superposition takes time $O(n^3)$.      
Similarly, the quantum algorithm for decomposing the simplicial complex
in terms of eigenvalues and eigenvectors of the combinatorial
Laplacian takes time $O(n^5)$, compared with $O(2^{2n})$ for 
classical algorithms.     The eigenvectors of the
kernels of the combinatorial Laplacian are related
to the representatives of the $k$th homology class via a boundary term.
How to extend the quantum algorithms given here
to construct the full barcode of persistent homology
and to construct the representatives
of the homology class directly is an open question. 
It would also be interesting to extend the quantum algorithmic methods
developed here to further algebraic and combinatorial problems,
e.g., Morse theory.

\vfill\noindent{\it Acknowledgments:} The authors give
heartfelt thanks to Mario Rasetti for suggesting the topic of topological
analysis of big data.  The authors acknowledge helpful conversations with 
Patrick Rebentrost, Barbara Terhal, and 
Francesco Vaccarino.  S.L. was supported by ARO, AFOSR, DARPA, and
Jeffrey Epstein.
P.Z. was supported by ARO MURI grant W911NF-11-1-0268 and by 
NSF grant PHY-969969. 

\vfil\eject

\noindent{\it References:}

\vskip 1cm

\smallskip\noindent [1]
A. Zomorodian, G. Carlsson, 
{\it Disc. Comput. Geom.}, {\bf 33}
249-274 (2005).

\smallskip\noindent [2]
V. Robins, 
{\it Top. Proc.} (1999),
503–532 (1999).

\smallskip\noindent [3]
P. Frosini and C. Landi, 
{\it Pattern Rec. 
Image Anal.} {\bf 9}, 596–603 (1999).

\smallskip\noindent [4] 
G. Carlsson, A. Zomorodian, A. Collins, and L. Guibas,  
{\it Proc. Sympos. Geom. Process.}, 127–138 (2004).

\smallskip\noindent [5]
 H. Edelsbrunner, D. Letscher, A. Zomorodian, 
{\it Discrete Comput. Geom.} {\bf 28}, 511–533 (2002).

\smallskip\noindent [6] M. Friedman, unpublished manuscript.

\smallskip\noindent [7] A. Zomorodian, {\it Computational Topology},
Algorithms and Theory of Computation Handbook, Second Edition, 
Chapter 3, section 2, 2009.

\smallskip\noindent [8]
F. Chazal, A. Lieutier, {\it Discrete Comput. Geom.} {\bf 37}(4),
601-617 (2007).


\smallskip\noindent [9]
D. Cohen-Steiner, H. Edelsbrunner, J. Harer, {\it Discrete
Comput. Geom.} {\bf 37}, 103-120 (2007).


\smallskip\noindent [10] S. Basu, 
{\it Discrete Comput. Geom.} {\bf 22}, 1-18 (1999);  
{\it Discrete Comput. Geom.} {\bf 30},  65-85 (2003);
{\it Found. Comput. Math.} {\bf 8}(1), 45-80 (2008);
Algorithms in real algebraic geometry: a survey,
arXiv:1409.1534, 2014.

\smallskip\noindent [11]
J. Friedman, Computing Betti numbers via combinatorial Laplacians. 
In Proc. 28th Ann. ACM Sympos. Theory Comput. (1996), pp. 386–391.


\smallskip\noindent 
[12] W.V.D. Hodge, {\it The Theory and Applications of Harmonic
Integrals,} CAmbridge University Press, Cambridge,
1941.

\smallskip\noindent
[13] J.R. Munkrees, {\it Elements of Algebraic Topology,}
Benjamin/Cummings, Redwood City CA, 1984.



\smallskip\noindent
[14] S. Butler, F. Chung,
{\it Ann. Comb.} {\bf 13}, 403-412 (2010).

\smallskip\noindent
[15] S. Maleti\'c, M. Rjkovi\'c,
{\it Eur. Phys. J. Spec. Top.} {\bf 212},
77-97 (2012).

\smallskip\noindent
[16] P. Niyogi, S. Smale, S. Weinberger,
{\it Siam J. Comput.} {\bf 40}, 646-663 (2011).

\smallskip\noindent 
[17] D. Kozlov, {\it Combinatorial Algebraic Topology. Algorithms 
and Computation in Mathematics, vol. 21.,}
Springer, Berlin (2008)

\smallskip\noindent
[18] R. Ghrist,  {\it Bull. Am. Math. Soc. (N.S.)},  {\bf 45(1)}, 
61–75 (2008).

\smallskip\noindent
[19] S. Harker, K. Mischaikow, M. Mrozek, V. Nanda, 
{\it Found. Comput. Math.} (2013); doi:10.1007/s10208-013-9145-0.

\smallskip\noindent
[20] K. Mischaikow, V. Nanda,
{\it Discrete Comput. Geom.}  {\bf 50,} 330–353 (2013).

\smallskip\noindent
[21] CHOMP: Computational homology project, http://chomp.rutgers.edu;

CAPD::RedHom: Reduction homology algorithms, http://redhom.ii.uj.edu.pl/.

\smallskip\noindent [22]
R. A. Servedio and S. J. Gortler,
{\it SIAM J. Comput.} {\bf 33}, 1067, (2004);
arXiv: quant-ph/0007036.

\smallskip\noindent [23]
A. Hentschel, B.C. Sanders, {\it Phys. Rev. Lett.}{ \bf 104} (2010),
063603; arXiv: 0910.0762.

\smallskip\noindent [24] H. Neven, V.S. Denchev, G.  Rose,
W.G. Macready,
Training a Large Scale Classifier with the Quantum Adiabatic Algorithm,
arXiv: quant-ph/0811.0416; arXiv: 0912.0779.

\smallskip\noindent [25] K.L. Pudenz, D.A. Lidar,
 {\it Quant. Inf. Proc. } {\bf 12}, 2027 (2013); arXiv: 1109.0325.

\smallskip\noindent [26] D. Anguita, S. Ridella, F. Rivieccion, R. Zunino,
{\it Neural Networks} {\bf 16}, 763-770 (2003).


\smallskip\noindent [28]
E. A\"imeur, G. Brassard, S. Gambs,
`Quantum speed-up for unsupervised learning,'
{\it Machine Learning} {\bf 90}, 261-287 (2013).

\smallskip\noindent [28] S. Lloyd, M. Mohseni, P. Rebentrost,
Quantum algorithms for supervised and unsupervised machine learning,
arXiv: 1307.0411.

\smallskip\noindent [29] P. Rebentrost, M. Mohseni, S. Lloyd,
`Quantum support vector machine for big feature and big data classification,'
 arXiv: 1307.0471.

\smallskip\noindent [30] S. Lloyd, M. Mohseni, P. Rebentrost,
`Quantum principal component analysis,' {\it Nat. Phys.} {\bf 3029},
 DOI: 10.1038 (2014); arXiv: 1307.0401.

\smallskip\noindent [31]
S. Aaronson, Quantum machine learning algorithms: read the fine print,

http://www.scottaaronson.com/papers/qml.pdf (2014).

\smallskip\noindent
[32] M. Kac, {\it Am. Math. Monthly} {\bf 73}, 1-23 (1966).

\smallskip\noindent [33] V. Giovannetti,
S. Lloyd, L. Maccone,  {\it Phys.Rev.Lett.} {\bf 100},
160501 (2008); arXiv: 0708.1879.

\smallskip\noindent [34] V. Giovannetti,
S. Lloyd, L. Maccone, {\it Phys.Rev.A} {\bf 78},
052310 (2008); arXiv: 0807.4994.

\smallskip\noindent [35] F. De Martini, V. Giovannetti, S. Lloyd, L. Maccone,
E. Nagali, L. Sansoni, F. Sciarrino, {\it Phys. Rev. A } {\bf 80},
010302(R) (2009); arXiv: 0902.0222.

\smallskip\noindent [36] M.S. Nielsen, I.L. Chuang, {\it Quantum
computation and quantum information}, Cambridge University Press,
Cambridge, 2000.

\smallskip\noindent
[37] K. Sadakane, N. Sugawara, T. Tokuyama, 
{\it Interdisc. Inf. Sci.} {\bf 8}, 129–136 (2002).

\smallskip\noindent
[38] A.W. Harrow, A. Hassidim, S. Lloyd,
{\it Phys. Rev. Lett.} {\bf 15}, 150502 (2009);
arXiv: 0811.3171.

\smallskip\noindent
[39] A. Yu. Kitaev, A.H. Shen, M.N. Vyalyi, {\it Classical and quantum 
computation}, Graduate Studies in Mathematics, Vol. 47, publications
of the American Mathematical Society, 2004.

\smallskip\noindent
[40] D.S. Abrams, S. Lloyd, {\it Phys. Rev. Lett.} {\bf 83}, 5162-5165
(1999);  arXiv: quant-ph/9807070.

\smallskip\noindent 
[41] P. Scheiblechner, {\it J. Complexity} {\bf 23}, 359-379 (2007).

\vfill\eject\end